\documentclass[aps,prc,twocolumn,preprintnumbers,amsmath,showpacs,floatfix,nofootinbib]{revtex4-1}
\usepackage{bm,graphicx}

\newcommand{\dd}{\mathrm{d}}

\begin{document}

\preprint{BI-TP 2013/025}

\title{Dissipative corrections to particle spectra and anisotropic flow from a saddle-point approximation to kinetic freeze out}

\author{Christian Lang} \email{chlang@physik.uni-bielefeld.de}
\author{Nicolas Borghini} \email{borghini@physik.uni-bielefeld.de}
\affiliation{Fakult\"at f\"ur Physik, Universit\"at Bielefeld, Postfach 100131, D-33501 Bielefeld, Germany}

\date{\today}

\begin{abstract}
  A significant fraction of the changes in momentum distributions induced by dissipative phenomena in the description of the fluid fireball created in ultrarelativistic heavy-ion collisions actually take place when the fluid turns into individual particles. 
  We study these corrections in the limit of a low freeze-out temperature of the flowing medium, and we show that they mostly affect particles with a higher velocity than the fluid. 
  For these, we derive relations between different flow harmonics, from which the functional form of the dissipative corrections could ultimately be reconstructed from experimental data. 
\end{abstract}


\maketitle

High-energy collisions of heavy nuclei, as performed at the Brookhaven Relativistic Heavy Ion Collider (RHIC) and the CERN Large Hadron Collider (LHC), lead to the formation of an extended fireball, the evolution of which is to a large degree well modelled by the laws of relativistic fluid dynamics (see e.g. Ref.~\cite{Huovinen:2013wma} for a recent review). 
Especially successful and promising for the extraction of precise values of the transport coefficients characterizing the created hot matter are the description of collective flow and in particular its anisotropies~\cite{Heinz:2013th}. 
The latter are usually quantified in terms of the Fourier harmonics $v_n(p_t,y)$ of the measured particle spectrum, which a priori depend on the particle type, transverse momentum $p_t$ and rapidity $y$. 

As is by now well established, these anisotropies in the final state momentum distributions are caused by asymmetries---so-called ``eccentricities''---in the initial-state geometry of the expanding matter. 
Simplifying the picture, one can identify three main sources for the flow coefficients in collisions of identical nuclei at ultrarelativistic energies. 
First, the approximate almond shape of the overlap region of the nuclei yields the major contribution to the elliptic flow $v_2$ in noncentral collisions~\cite{Ollitrault:1992bk}.
Secondly, the event-by-event fluctuations in the positions of nucleons---or more generally of the colliding degrees of freedom---inside the nuclei at the time of the collision lead to deviations of the geometry from the smooth shape corresponding to the overlap of ideal spheres. 
These initial-state fluctuations give rise to triangular flow $v_3$~\cite{Alver:2010gr} and a rapidity-even (and thereby present at midrapidity) contribution to directed flow $v_1$~\cite{Teaney:2010vd}, as well as to the elliptic flow measured in most central collisions. 
They also contribute to the ``higher harmonics'' $v_4$, $v_5$, $v_6$\ldots, which are, however, also to a large extent controlled by a third phenomenon, namely the mixing of sizable lower harmonics: $v_4$ contains a large nonlinear contribution from $v_2^2$~\cite{Borghini:2005kd}; $v_5$, from the product $v_2v_3$~\cite{Gardim:2011xv,Teaney:2012ke}; or $v_6$, from $v_2^3$ and $v_3^2$~\cite{Bravina:2013ora}. 

The evolution from the initial geometry to the final-state anisotropies may be viewed as a filtering process, in which 
the first three flow harmonics respond linearly to corresponding initial-state asymmetries, while the higher harmonics constitute some  nonlinear response. 
The filter characteristics, as e.g.\ the proportionality coefficients in the linear-response regime, reflect the properties of the expanding medium. 
In particular, if the fireball is modelled as a fluid, its dissipative features---like shear and bulk viscosity or the relaxation time of the viscous tensor---govern the response. 
Relating the flow harmonics to the initial eccentricities gives then in principle access to the fluid transport coefficients, which is one of the goals of present heavy-ion physics. 

In a hydrodynamical approach, dissipative phenomena enter the description in a twofold way. 
They first play a role all along the evolution, which is mathematically accounted for by the fact that the fluid velocity obeys equations of dissipative relativistic hydrodynamics, namely Navier--Stokes or second-order equations~\cite{Huovinen:2013wma}. 
Strictly speaking, this necessitates knowledge of the temperature dependence of the transport coefficients over the range covered over the fireball history. 
Dissipation also affects the endpoint of the fluid evolution, that is the transition from a continuous medium to a collection of particles. 
This corresponds in so-called ``hybrid models'' to the switch from hydrodynamics to a particle transport model~\cite{Huovinen:2012is}, or in a more simplified picture, which we shall hereafter adopt, to the sudden (kinetic) freeze out of the fluid into noninteracting particles. 
Modelling this juncture with the Cooper--Frye prescription~\cite{Cooper:1974mv}, 
the invariant distribution of particles decoupling from of a fluid with four-velocity $u^\mu(\mathsf{x})$ reads%
\footnote{Here and throughout the paper, we use a metric with $(+,-,-,-)$ signature and denote four-vectors in sans serif font and three-vectors in boldface.}
\begin{equation}
\label{Cooper-Frye}
E_{\textbf{p}}\frac{\dd^3N}{\dd^3\textbf{p}} = 
  \frac{g}{(2\pi)^3}\!\int_\Sigma\!f\bigg(\frac{\mathsf{p}\cdot\mathsf{u}(\mathsf{x})}{T}\bigg)_{} p^\mu\dd^3\sigma_\mu(\mathsf{x}),
\end{equation}
with $\Sigma$ the freeze-out hypersurface, here defined by a constant temperature $T$, and $g$ the degeneracy factor for the particles. 
$f$ denotes a phase space distribution, the precise form of which depends on the particle type---boson or fermion---and on the dissipative properties of the fluid. 
Thus, for the decoupling from a perfect fluid, $f$ is given by the equilibrium thermal distribution---Bose--Einstein or Fermi--Dirac, although we shall from now on focus on the regime where quantum statistics effects are negligible and approximate either of them by the Maxwell--Boltzmann distribution, which will be denoted $f_0$. 
If the freezing-out fluid is dissipative, $f$ contains extra terms, to ensure the continuity of the energy-momentum tensor at decoupling. 
These corrections have been computed, in the case of a transition to an ideal single-component Boltzmann gas, for a fluid with finite shear~\cite{Teaney:2003kp} or bulk viscosity~\cite{Dusling:2007gi}, or a conformal fluid obeying second-order dissipative hydrodynamics~\cite{Teaney:2013gca}. 
It has, however, been recognised that more realistic corrections are needed---and some have been computed in various models~\cite{Denicol:2009am,Monnai:2009ad,Dusling:2009df,Pratt:2010jt,Molnar:2011kx,Dusling:2011fd}---and there have been attempts to constrain them from the available experimental data~\cite{Luzum:2010ad}. 

In the present study, we wish to pursue this avenue and investigate whether the functional form of the dissipative corrections to the phase space distribution at the end of the hydrodynamic evolution, in particular their dependence on the emitted particle momentum, can be reconstructed from the shape of the flow harmonics. 
For that purpose, we follow the idea of Ref.~\cite{Borghini:2005kd} and compute the Cooper--Frye integral~(\ref{Cooper-Frye}) within a saddle-point approximation (Sect.~\ref{s:saddle-point_app}). 
This leads us to identify two main classes of particles, ``slow'' and ``fast'', according to how their velocity compares to the maximal velocity of the fluid flowing in the direction of their momentum. 
We show in Sect.~\ref{s:slow-particles} that for slow particles, the dissipative effects coming from freeze out are actually minimal, so that the qualitative behaviours found in the ideal case remain valid. 
Turning then to fast particles (Sect.~\ref{s:fast-particles}), we investigate the dissipative corrections from freeze out on anisotropic flow and find that by using relations between different flow harmonics, it may be possible to constrain the functional form of these effects from the data. 
Eventually, in Sect.~\ref{s:Discussion} we summarise our findings and compare some of our results to ``exact'' numerical computations of the Cooper--Frye integral for a toy freeze-out profile, so as to gauge the validity of the saddle-point approximation. 

Throughout this paper, we leave aside fluctuations, i.e.\ we work with exactly reconstructed flow harmonics $v_n$, not with their root mean squares or other similar quantities as extracted from various analysis methods.

\section{Saddle-point computation of the Cooper--Frye integral}
\label{s:saddle-point_app}

To investigate the effect of dissipative corrections due to the matching between fluid and particles, we shall not assume a specific flow profile (like e.g.\ Bjorken flow or a blast wave) as was done in previous analytical studies. 
Instead, we bypass knowledge of the freeze-out hypersurface in the Cooper--Frye prescription by approximating the integral with the saddle-point method. 
Quite naturally, the trade off for this approximation is a restriction of the range of validity of our results, which will only hold in given transverse momentum intervals, and for some observables only. 

In most models analysed so far, with the exception of Ref.~\cite{Pratt:2010jt}, the single-particle phase space distribution at decoupling is taken to be of the form 
\begin{equation}
\label{f_vs_f0_gen}
f(\mathsf{x}, \mathsf{p}) = \big[1 + \overline{\delta f}(\mathsf{x}, \mathsf{p})\big]
  f_0\bigg(\frac{\mathsf{p}\cdot\mathsf{u}(\mathsf{x})}{T}\bigg).
\end{equation}
That is, dissipative effects contribute an additive term proportional to the equilibrium distribution---in addition to the modification of the flow velocity profile $\textsf{u}(\textsf{x})$. 
For the sake of consistency of the hydrodynamic description, the modulus of the ``reduced'' correction $\overline{\delta f}$ should be (much) smaller than 1. 
Here, we shall also adopt the ansatz~(\ref{f_vs_f0_gen}), and further use the condition $|\overline{\delta f}|\ll 1$ to replace the actual saddle point of the integrand in Eq.~(\ref{Cooper-Frye}), corresponding to $f$, by the saddle point obtained with $f_0$ only. 
It can easily be checked that the changes introduced by this simplification are actually of second order in the small parameters controlling $\overline{\delta f}$. 
Since we consider the regime of not too small momenta where $f_0$ is given by the Maxwell--Boltzmann distribution, the saddle point is then the point(s) on the freeze-out hypersurface where $\mathsf{p}\cdot\mathsf{u}(\mathsf{x})/T$ is minimum. 
As this was already studied in Ref.~\cite{Borghini:2005kd}, we shall in the main body of the text only review the findings, relegating more detailed calculations to Appendix~\ref{app:calculations}. 

For longitudinal motion, the saddle point selects regions of the freezing-out fluid with the same rapidity $y_{\rm f}$ as that ($y$) of the emitted particles. 
Regarding transverse motion, a particle with azimuthal angle (with respect to a given reference) $\varphi$ is actually stemming from a fluid cell with transverse velocity $\mathbf{u}_t$ pointing along $\varphi$, i.e.\ parallel to the particle transverse momentum  $\mathbf{p}_t$.

To further specify the transverse velocity of the fluid corresponding to particles with a given transverse momentum, one needs to introduce the maximum value of $|\mathbf{u}_t|$ for fixed rapidity and azimuth, $u_{\max}(y,\varphi)$. 
A typical value for $u_{\max}$ at midrapidity in heavy-ion collisions at maximum RHIC energy or at the LHC is about 1. 
``Slow'' resp.\ ``fast'' particles are then defined as those with a transverse velocity $p_t/m$ smaller resp.\ larger than $u_{\max}(y,\varphi)$. 
The former are emitted by a fluid region with respect to which they are at rest, i.e.\ such that $\mathbf{u}_t = \mathbf{p}_t/m$; 
one then finds at once $\mathsf{u}(\mathsf{x}_{\mathrm{s.p.}}) = \mathsf{p}/m$ at the saddle point $\mathsf{x}_{\mathrm{s.p.}}$, which also gives 
\begin{equation}
\label{p.u_slow}
\mathsf{p} \cdot \mathsf{u}(\mathsf{x}_{\mathrm{s.p.}}) = m\quad\text{for slow particles}. 
\end{equation}
We shall use this result in next section. 

For fast particles, the minimum of $\mathsf{p} \cdot \mathsf{u}(\mathsf{x})$ is reached at a saddle point $\mathsf{x}_{\mathrm{s.p.}}$ where the fluid transverse velocity reaches its maximum $u_{\max}(y,\varphi)$, and some straightforward algebra yields
\begin{align}
\label{p.u_fast}
\mathsf{p} \cdot \mathsf{u}(\mathsf{x}_{\mathrm{s.p.}}) = m_t u_{\max}^0(y,\varphi) &- p_t u_{\max}(y,\varphi) \cr
 &\text{ for fast particles},
\end{align}
where we have defined $u_{\max}^0(y,\varphi) \equiv \sqrt{1+u_{\max}(y,\varphi)^2}$, while $m_t\equiv\sqrt{m^2+p_t^2}$ is the usual ``transverse mass''.

Before we exploit Eqs.~(\ref{p.u_slow}) and (\ref{p.u_fast}), let us recall that slow or fast particles must actually obey more stringent conditions for the saddle-point approximation to hold. 
Thus, slow particles should have a mass significantly larger than the freeze-out temperature, which unfortunately excludes pions and might only marginally be fulfilled by kaons. 
In turn, fast particles should obey condition~(\ref{criterion_fast}) from Appendix~\ref{app:calculations}, which translates into a species-dependent lower bound on the particle transverse momentum. 
In either case, the smaller the freeze-out temperature is, the better the saddle-point approximation is.

\section{Slow particles}
\label{s:slow-particles}

For slow particles decoupling from an ideal fluid, it was found that Eq.~(\ref{p.u_slow}) leads to the remarkable property that the particle distribution resulting from the Cooper--Frye prescription is simply (the degeneracy factor times) a function of mass multiplying a species-independent function of the particle transverse velocity $p_t/m$, azimuth $\varphi$, and rapidity $y$~\cite{Borghini:2005kd}:
\begin{equation}
\label{slow:scaling_law}
E_{\textbf{p}}\frac{\dd^3N}{\dd^3\textbf{p}} = c(m)\,F\bigg(\frac{p_t}{m}, \varphi, y\bigg).
\end{equation}
As a consequence, the particle spectra for different species, plotted vs.\ $p_t/m$ at a given rapidity, should only differ by a normalisation factor. 
Expanding the particle distribution in Fourier series of the azimuthal angle, the Fourier coefficients $v_n(p_t/m, y)$ should be identical for different species of slow particles. 
Plotting as a function of transverse momentum $p_t$, instead of transverse velocity, one finds the so-called ``mass ordering'' of the flow coefficients, with $v_n(p_t)$ being smaller for heavier particles---thanks to the fact that $v_n$ is a monotonously increasing function of transverse momentum. 

As we shall show next, these generic features---namely transverse momentum spectra as product of a particle type dependent coefficient and a universal function of the particle velocity and anisotropic flow coefficients depending only on $p_t/m$ and $y$---actually persist for slow particles decoupling from a dissipative fluid, at least as far as first-order or conformal second-order effects are concerned.
Note, however, that the prefactors $c(m)$ and the shape of the species-independent function $F$ do depend on the form of the dissipative corrections. 

To see that the latter still lead to a functional dependence of the type~(\ref{slow:scaling_law}), we have to inspect the form of the dissipative corrections at freeze out more closely. 

Consider first the correction accounting for shear viscous effects. 
This contribution contains at least a multiplicative factor $\pi^{\mu\nu}_{\rm shear}(\mathsf{x})_{}p_\mu p_\nu$, with $\pi^{\mu\nu}_{\rm shear}$ the shear stress tensor. 
For our discussion, the latter possesses the important property that it is orthogonal to the fluid velocity, $\pi^{\mu\nu}_{\rm shear}u_\mu=0$. 
As we have seen above, the saddle point for slow particles is such that $\mathsf{u}(\mathsf{x}_{\mathrm{s.p.}}) = \mathsf{p}/m$, which yields at once 
\[
\pi^{\mu\nu}_{\rm shear}(\mathsf{x}_{\mathrm{s.p.}})_{}p_\mu p_\nu \propto
  \pi^{\mu\nu}_{\rm shear}(\mathsf{x}_{\mathrm{s.p.}})_{}u_\mu(\mathsf{x}_{\mathrm{s.p.}})_{} u_\nu(\mathsf{x}_{\mathrm{s.p.}}) = 0.
\]
Thus, the additive correction at decoupling from shear viscosity vanishes for slow particles in the saddle-point approximation. 

The bulk viscous term is also readily dealt with. 
Quite generally, it should be of the form 
\[
\overline{\delta f}^{(1)}_{\mathrm{bulk}} = 
  C_{\mathrm{bulk}}\big(\mathsf{p}\cdot\mathsf{u}(\mathsf{x}),\mathsf{p}^2\big)_{}\Pi(\mathsf{x}),
\]
with $\Pi(\mathsf{x}) = \zeta\,\partial_\mu u^\mu(\mathsf{x})$ the bulk pressure and $C_{\mathrm{bulk}}$ a function. 
With the help of Eq.~(\ref{p.u_slow}) valid for slow particles, one sees that the arguments are actually simply $m$ and $m^2$, i.e.\ momentum independent. 
In turn, the bulk pressure at freeze out only includes the expansion rate $\partial_\mu u^\mu$, taken at the same (saddle) point for particles having the same transverse velocity. 
Again, one finds that the particle distribution depends on momentum only through the variables $p_t/m$, $y$, and $\varphi$, so that the conclusions found for the freeze out from an ideal fluid remain valid, albeit with modified factors $c(m)$ and shape $F(p_t/m,\varphi, y)$.  

Conformal second-order corrections to the phase space distribution of slow particles can be handled in the same way as the shear or bulk viscous terms above.  
Consider thus Eq.~(\ref{df2_Teaney-Yan}) in appendix~\ref{app:Teaney-Yan_eqs}, in which the corrections computed in Ref.~\cite{Teaney:2013gca} are repeated.
The five first terms contain contractions of the particle four-momentum with tensors orthogonal to the fluid four-velocity, and they thus yield a vanishing contribution at the saddle points for slow particles, by the same argument as for first-order shear corrections. 
The last term of Eq.~(\ref{df2_Teaney-Yan}) involves on the one hand the quantity $\bar{\xi}_{4p}$ defined in Eq.~(\ref{xi4}), which in turn only depends on the energy and momentum of the particle in the fluid local rest frame: at the saddle point, these are simply the particle mass and zero, respectively.
Besides, that last term also involves the (second-order) dissipative stress tensor, which has to be evaluated at the same ``universal'' saddle point for particles having a given transverse velocity, and thus will contribute a term depending on momentum only through $p_t/m$, $y$, and $\varphi$. 
All in all, the correction will again lead to a distribution obeying the scaling law~(\ref{slow:scaling_law}). 

One can anticipate that nonconformal second-order corrections can be dealt with as easily. 
Yet it may be noted that freeze out is most commonly assumed to take place at a temperature at which the fluid, according to lattice gauge field theory results, is approximately conformal, so that such corrections might actually turn out to be quite small, especially for particles that freeze out later.

\section{Fast particles}
\label{s:fast-particles}

Let us now turn to fast particles. 
Inserting Eq.~(\ref{p.u_fast}) in the integrand of the Cooper--Frye formula, one deduces 
\begin{equation}
\label{spectrum_fast}
E_{\textbf{p}}\frac{\dd^3N}{\dd^3\textbf{p}} \propto \exp\!\bigg[\frac{p_t u_{\max}(y,\varphi) - m_t u_{\max}^0(y,\varphi)}{T} \bigg]. 
\end{equation}
The omitted prefactor depends on the dissipative corrections, estimated at the saddle point, as well as on the behaviour of the velocity in the neighbourhood of the saddle point, which necessitates more detailed knowledge on the flow profile at freeze out. 
To bypass the need for this knowledge, we shall focus on the azimuthal anisotropies of the particle distribution, i.e.\ the flow coefficients $v_n$, which do not depend on the absolute normalisation of the spectrum. 
For the sake of brevity, we shall from now on drop the rapidity $y$ from our expressions. 

We introduce the expansion of the maximum transverse flow velocity $u_{\max}(y,\varphi)$ at freeze out as a Fourier series 
\begin{equation}
\label{u_max(phi)_expansion}
u_{\max}(\varphi) = \bar{u}_{\max} \bigg[ 1 + 2\sum_{n\geq 1}V_n\cos n(\varphi-\Psi_n)\bigg],
\end{equation}
with $\Psi_n$ the $n$-th harmonic symmetry-plane angle. 
Given any realistic velocity profile, $\bar{u}_{\max}$ and the anisotropies $V_n$---which naturally all depend on $y$---are easily reconstructed. 
The three-velocity value corresponding to the average maximum transverse flow velocity $\bar{u}_{\max}$ will be denoted
\begin{equation}
\label{def_v_max}
\bar{v}_{\max}\equiv\frac{\bar{u}_{\max}}{\sqrt{1+\bar{u}_{\max}^2}}.
\end{equation}
A typical value of 1 for $\bar{u}_{\max}$ amounts to $\bar{v}_{\max}\simeq 0.7$. 
In turn, the Fourier coefficients $V_n$ are assumed to be small, say of order 0.05 or smaller. 
Hereafter, we shall assume that they obey the hierarchy $V_2\gtrsim V_3 \gg V_1, V_4, V_5$, and that higher coefficients vanish. 
Yet our calculations can easily be repeated with any other hierarchy of the anisotropies of  the maximum transverse flow velocity at freeze out. 

Expansion~(\ref{u_max(phi)_expansion}) is reported in Eq.~(\ref{spectrum_fast}), namely into the exponent and---if necessary---in the prefactor. 
In the latter, one should strictly speaking know the Fourier expansions of various combinations of the derivatives of the flow velocity $\mathsf{u}(\mathsf{x})$ around the saddle point---for instance, the azimuthal dependence of the components of the shear stress tensor. 
We shall for simplicity neglect this dependence, considering that it only represents a small modulation of a quantity which is already small in itself. 
There is however no difficulty of principle in including this refinement, at the cost of introducing new Fourier coefficients for each azimuthally dependent quantity. 

Some straightforward algebra involving the Taylor expansion of the exponent in Eq.~(\ref{spectrum_fast}) then yields the Fourier coefficients of the invariant single-particle distribution for fast particles. 
Restricting ourselves to the first five harmonics, one finds
\begin{subequations}
\label{vn(pt)}
\begin{align}
v_1(p_t) &= \big[\mathcal{I}(p_t)-\mathcal{D}(p_t)\big]V_1 \cr 
 &\qquad\qquad+ \big[\mathcal{I}(p_t)^2-\mathcal{I}(p_t)_{}\mathcal{D}(p_t)\big]V_2V_3, \label{v1(pt)}\\
v_2(p_t) &= \big[\mathcal{I}(p_t)-\mathcal{D}(p_t)\big]V_2, \label{v2(pt)}\\
v_3(p_t) &= \big[\mathcal{I}(p_t)-\mathcal{D}(p_t)\big]V_3 + \mathcal{O}(V_1V_2), \label{v3(pt)}\\
v_4(p_t) &= \big[\mathcal{I}(p_t)-\mathcal{D}(p_t)\big]V_4 \cr 
 &\qquad\qquad+ \bigg[\frac{\mathcal{I}(p_t)^2}{2}-\mathcal{I}(p_t)_{}\mathcal{D}(p_t)\bigg]V_2^2, \label{v4(pt)}\\
v_5(p_t) &= \big[\mathcal{I}(p_t)-\mathcal{D}(p_t)\big]V_5 \cr 
 &\qquad\qquad+ \big[\mathcal{I}(p_t)^2-\mathcal{I}(p_t)_{}\mathcal{D}(p_t)\big]V_2V_3. \label{v5(pt)}
\end{align}
\end{subequations}
In these relations, $\mathcal{I}(p_t)$ is a simple function that does not depend on the dissipative corrections to the single-particle phase space distribution, namely
\begin{equation}
\label{I(pt)}
\mathcal{I}(p_t) = \frac{\bar{u}_{\max}}{T}(p_t -m_t\bar{v}_{\max}). 
\end{equation}
For fast particles, $p_t /m_t > \bar{v}_{\max}$ so that $\mathcal{I}(p_t)$ is always positive. 
The function $\mathcal{D}(p_t)$ represents the term stemming from the dissipative contributions $\overline{\delta f}$ to the phase space distributions, and it vanishes when these are absent, that is, for particles freezing out from an ideal fluid. 
The actual form of $\mathcal{D}(p_t)$, in particular the functional dependence on $p_t$, directly reflects that of $\overline{\delta f}$. 
We give as an example the function $\mathcal{D}(p_t)$ resulting from considering only first-order shear viscous corrections as given by Grad's prescription in appendix~\ref{app:df1_shear}.
More generally, one can compute $\mathcal{D}(p_t)$ starting from any ansatz for $\overline{\delta f}$. 
This can then be compared with the shape constrained from experimental results as we explain below. 

Before that, let us discuss the relations~(\ref{vn(pt)}), starting with the ``ideal case'' when dissipative effects vanish, i.e.\ $\mathcal{D}(p_t) = 0$. 
Equations~(\ref{v2(pt)}) and (\ref{I(pt)}) then reduce to Eq.~(8) of Ref.~\cite{Borghini:2005kd} for $v_2(p_t)$. 
Similarly, one recovers the nonlinear ideal relations $v_4(p_t) \simeq v_2(p_t)^2/2$~\cite{Borghini:2005kd} and $v_5(p_t) \simeq v_2(p_t)_{}v_3(p_t)$~\cite{Teaney:2012ke}, valid in the large $p_t$ regime where the linear contributions to these higher harmonics become negligible. 

For particles freezing out of a dissipative fluid, $\mathcal{D}(p_t)$ is now non-zero. 
Another change, which is not reflected in our notations, affects the average $\bar{u}_{\max}$ and Fourier coefficients $V_n$ of the maximum flow velocity at freeze out, which do depend on the amount of dissipation along the system evolution. 
In the following discussion, we take the values of these quantities as fixed, and we only consider the effect of including $\mathcal{D}(p_t)$ or not. 

First, one sees at once that when $\mathcal{D}(p_t)>0$, its inclusion leads to a decrease of every flow harmonic~(\ref{vn(pt)}). 
Now, the actual sign of $\mathcal{D}(p_t)$ depends on the flow profile at freeze out. 
In existing hydrodynamical simulations, it has turned out to be positive, as hinted at in particular by the decrease of $v_2(p_t)$ at large transverse momentum and midrapidity, which a posteriori explains our choice of signs in Eqs.~(\ref{vn(pt)}).  
There are, however, theoretical grounds to expect that the bulk viscous contribution to $\mathcal{D}(p_t)$ could change sign~\cite{Luzum:2010ad}---although it is not clear whether this should happen into the fast particle region---, so that probably no definite statement can be made. 

Among the relations~(\ref{vn(pt)}), some show obvious similarities. 
Thus, Eqs.~(\ref{v2(pt)}) and (\ref{v3(pt)}) predict a constant ratio $v_3(p_t)/v_2(p_t)$ in case the hierarchy $V_3\sim V_2\gg V_1$ holds.\footnote{We checked for such a regularity in the ALICE data for identified particles in semi-peripheral Pb--Pb collisions and found that the ratio $v_3(p_t)/v_2(p_t)$ for kaons and (anti)protons is identical in the transverse momentum range where they are ``fast''; the ratio is however not constant, but increasing. 
  This might be due to the fact that in the considered centrality range, the hierarchy of flow harmonics does not hold, so that the nonlinear $V_1V_2$ contribution to $v_3(p_t)$ starts playing a role. 
  In the absence of the relevant $v_1(p_t)$ data, we could not investigate this idea further.}
Likewise, Eqs.~(\ref{v1(pt)}) and (\ref{v5(pt)}) are very similar and predict analogous $v_1(p_t)$ and $v_5(p_t)$ in the regime where the linear contributions to these harmonics become negligible with respect to the $V_2V_3$ term. 
Let us emphasise that these similarities between different flow harmonics hold in the regime of fast particles, i.e.\ far from $\mathbf{p}_t=\mathbf{0}$, where the analyticity of the momentum distribution induces different scaling behaviours for each flow harmonic~\cite{Danielewicz:1994nb}. 

Another finding from Eqs.~(\ref{vn(pt)}) is that the nonlinear relations valid in the ideal case no longer hold. 
Thus, $v_4(p_t)/v_2(p_t)^2$ is now smaller than $\frac{1}{2}$ when $V_4$ can be neglected and, more generally, this ratio is decreased by the inclusion of the dissipative correction at freeze out $\mathcal{D}(p_t)$, whether or not $V_4$ is taken into account.
In contrast, when neglecting $V_5$ the ratio $v_5(p_t) / v_2(p_t)_{}v_3(p_t)$ increases for $\mathcal{D}(p_t)\neq 0$ and is thus larger than the ``ideal'' value of 1. 
These qualitative results are borne out by results either from a Boltzmann transport model~\cite{Gombeaud:2009ye} or from hydrodynamical simulations~\cite{Luzum:2010ae,Teaney:2012ke}.

The nonlinear relations can actually be exploited for more quantitative results, still in the case of negligible linear contributions. 
Thus, one finds from Eqs.~(\ref{v2(pt)}), (\ref{v3(pt)}) and (\ref{v5(pt)})
\begin{equation}
\label{vn_relation_1}
\frac{v_5(p_t)-v_2(p_t)_{}v_3(p_t)}{v_3(p_t)} = \mathcal{D}(p_t)_{}V_2,
\end{equation}
or similarly, using Eqs.~(\ref{v2(pt)})--(\ref{v4(pt)})
\begin{equation}
\label{vn_relation_2}
v_2(p_t)^2 - 2v_4(p_t)= \mathcal{D}(p_t)^2V_2^2. 
\end{equation}
That is, one can isolate the dissipative contribution from decoupling to $v_2(p_t)$---and more generally, the term $\mathcal{D}(p_t)$. 
Here we gave two independent relations from which the dissipative term can be experimentally constrained and then compared with the functional form derived from a functional ansatz for $\overline{\delta f}$. 

\section{Discussion}
\label{s:Discussion}

In the previous two sections, we investigated the effect of the dissipative correction $\overline{\delta f}$ to the phase space distribution of particles at freeze out on the particle spectrum. 

We first found that for slow particles, which are emitted by a fluid region moving at the same velocity, the results valid in the ideal case are qualitatively not modified by dissipative effects: spectra for different particles coincide, up to a multiplicative factor, when considered as a function of transverse velocity $p_t/m$; and this implies mass ordering of the flow harmonics. 
This result starts bridging the gap between the limiting cases of ideal fluid dynamics on the one side~\cite{Borghini:2005kd}, and Boltzmann transport calculations with very few scatterings per particle on the other side~\cite{Borghini:2010hy}, in which the role of velocity as scaling variable was emphasised. 

For fast particles, we focussed on the anisotropic flow coefficients $v_n(p_t)$. 
Here, we recovered the qualitative behaviours already identified in numerical simulations for both ``linearly'' and ``nonlinearly responding'' harmonics. 
In addition, we showed that by investigating appropriate combinations of several flow harmonics---involving a nonlinear one and the linear ones that contribute to it---, one could ideally reconstruct from the data the momentum dependence of $\overline{\delta f}$. 
We only gave two examples~(\ref{vn_relation_1}) or~(\ref{vn_relation_2}) of such relations, but other can be derived, in particular by generalizing our present ``single-particle'' study to multiparticle correlations.  

To gauge the validity of our results, especially of the relations found for fast particles, we tested them on the flow coefficients arising from the numerical integration over a three-dimensional freeze-out hypersurface $\Sigma$ of some flow profile. 
More precisely, we took for $\Sigma$ an infinite (along the longitudinal axis) azimuthally symmetric cylinder of radius $R$, at a constant proper time $\tau_{\textrm{f.o.}}$. 
Using as space-time coordinates the proper time $\tau$, cylindrical coordinates $r,\phi$ and space-time rapidity $\eta_s$, we assumed for the fluid velocity on $\Sigma$ a generalised blast wave-like profile for the radial coordinate~\cite{Siemens:1978pb,Huovinen:2001cy}
\begin{equation}
\label{blastwave}
u^r(r,\phi) = \bar{u}_{\max}\frac{r}{R} \bigg( 1 + 2\sum_{n=1}^5V_n\cos n\phi\bigg),
\end{equation}
together with $u^\phi = u^\eta = 0$ in the azimuthal and $\eta_s$ directions, as well as naturally $u^\tau=\sqrt{1+(u^r)^2}$ in the timelike direction. 
From this expression, one directly reads off the maximal transverse velocity [cf. Eq.~(\ref{u_max(phi)_expansion})].
The plots presented below were obtained with the values $R=7.5$~fm, $\tau_{\textrm{f.o.}}=5.25$~fm/$c$, $T_0=160$~MeV, $\bar{u}_{\max}=0.55$ and $V_2=0.05$---corresponding to the choice made in Ref.~\cite{Teaney:2003kp}---and additionally $V_3=0.05$, and all other $V_n=0$. 
We performed tests with other values, without changing the findings we now report. 

With such a choice of flow profile, 7 out of the 10 different components of the shear viscous stress tensor $\pi^{\mu\nu}_{\textrm{shear}}$ are non-zero. 
Nevertheless, in our saddle-point approximation we only keep $\pi^{rr}_{\textrm{shear}}$ as explained in appendix~\ref{app:df1_shear}. 
With the relatively small chosen value of $\bar{u}_{\max}$ and with a ratio $\eta/s=0.16$, the coefficient $C'_{\mathrm{shear}}$ defined in that appendix is of order 0.6, which ensures that some of the terms we have neglected in deriving the correction term~(\ref{D(pt)_shear}) remain small as long as $p_t$ (or more accurately $p_t-m_t\bar{v}_{\max}$) is not too large. 
One can naturally depart from this assumption, at the cost of considering a more lengthy formula for the correction ${\cal D}(p_t)$. 

Given this set up for our numerical toy model for the Cooper--Frye distribution, we can compare its results with the findings within the saddle-point approximation, focussing on fast particles---that is, on the region $p_t\gtrsim 0.6$~GeV/$c$ for pions. 
To begin with a blunt statement, the saddle-point behaviours~(\ref{vn(pt)}) represent a bad approximation to those of the numerical simulation in the $p_t$ range which seems reasonable for the comparison to experimental data. 
To list a few discrepancies, which already appear for the decoupling from an ideal fluid: 
$v_2(p_t)$ in the exact blast wave model grows quadratically at low $p_t$ (for pions, until about 1--1.5~GeV/$c$), while Eq.~(\ref{v2(pt)}) is almost linear. 
Reltions~(\ref{v2(pt)}) and~(\ref{v3(pt)}) predict parallel behaviours for $v_2(p_t)$ and $v_3(p_t)$---with our choice of values for $V_2$ and $V_3$, they should be equal---, while the full computation gives $v_3(p_T)$ significantly smaller [the ratio $v_3(p_t)/v_2(p_t)$ for pions grows from 0.5 at 1~GeV/$c$ to 0.8 at 3~GeV/$c$]. 
Below 3~GeV/$c$, the ``exact'' (we shall from now on use this short formulation) $v_1(p_t)$ and $v_4(p_t)$ differ by more than a factor 2, while Eqs.~(\ref{v1(pt)}) and (\ref{v4(pt)}) give them equal; on the other hand, the exact $v_4(p_t)$ almost equals $v_5(p_t)$, while relations~(\ref{vn(pt)}) predict a factor 2 in the ideal case. 
In short, the approximations~(\ref{vn(pt)}) are quite unsatisfactory below 3~GeV/$c$. 
Let us, however, note that they become much better above 5~GeV/$c$, as was actually already observed for the nonlinear relations between higher harmonics and the lower ones in realistic hydrodynamical computations~\cite{Teaney:2012ke}. 
This region is probably not relevant for comparison to experimental data, but might help with the understanding of numerical fluid dynamics simulations.

\begin{figure}[t!]
\includegraphics*[width=\linewidth]{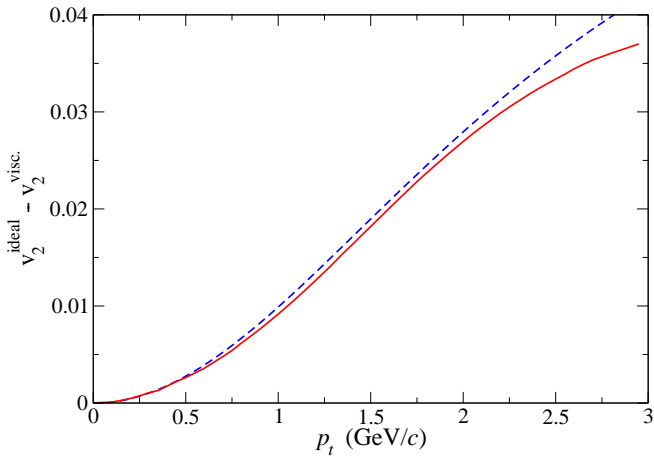}
\caption{\label{fig:delta_v2} Difference between the values of the pion elliptic flow $v_2(p_t)$ in the ideal and shear viscous cases, for a full computation of the Cooper--Frye integral (dashed) or within the saddle-point approximation (full curve).}
\end{figure}
Despite our having just criticised the ``absolute'' predictions~(\ref{vn(pt)}), we shall now argue that the saddle-point approximation captures the effect of dissipative effects at freeze out in an astonishingly good manner. 
To illustrate this point, we display in Fig.~\ref{fig:delta_v2} the difference between the ideal and shear viscous $v_2(p_t)$---computed with the same values of all parameters listed below Eq.~(\ref{blastwave}), in particular $\bar{u}_{\max}$---as given by the exact numerical integration of the Cooper--Frye integral (dashed curve).
This difference should only reflect the effect of the dissipative correction $\overline{\delta f}$, which within our saddle-point calculation, represented by the full curve, is simply $\mathcal{D}(p_t)_{}V_2$, with $\mathcal{D}(p_t)$ given by Eq.~(\ref{D(pt)_shear}). 
The agreement between the numerical and analytical results is obviously excellent, especially when keeping in mind that the $v_2(p_t)$ values themselves are quite poorly approximated by the saddle-point calculation. 
Note that the growing departure between the two curves above 2~GeV/$c$ arises from our having discarded terms in deriving the simple formula~(\ref{D(pt)_shear}) and can be cured by including more terms. 
On the other hand, we have no explanation for the excellent agreement at low momentum, outside the regime of fast particles. 

\begin{figure}[t!]
\includegraphics*[width=\linewidth]{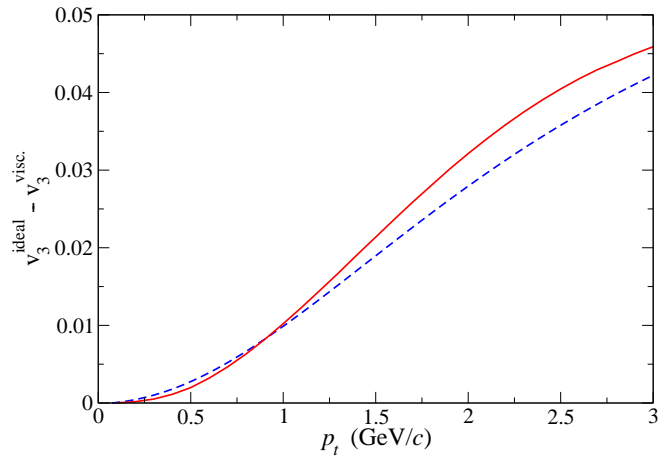}
\caption{\label{fig:delta_v3} Same as Fig.~\ref{fig:delta_v2} for triangular flow $v_3(p_t)$.}
\end{figure}
As mentioned above, the exact $v_2(p_t)$ and $v_3(p_t)$ are quite different from each other, in contrast to the saddle-point prediction. 
In Fig.~\ref{fig:delta_v3}, we show the difference of the ideal and viscous $v_3(p_t)$, analogous to Fig.~\ref{fig:delta_v2}. 
Again, the saddle-point result $\mathcal{D}(p_t)_{}V_3$ provides a good approximation to the full computation, especially given that the saddle-point calculation for either of the $v_3(p_t)$ is too large by a factor of about two or more across this $p_t$ range. 

\begin{figure}[t!]
\includegraphics*[width=\linewidth]{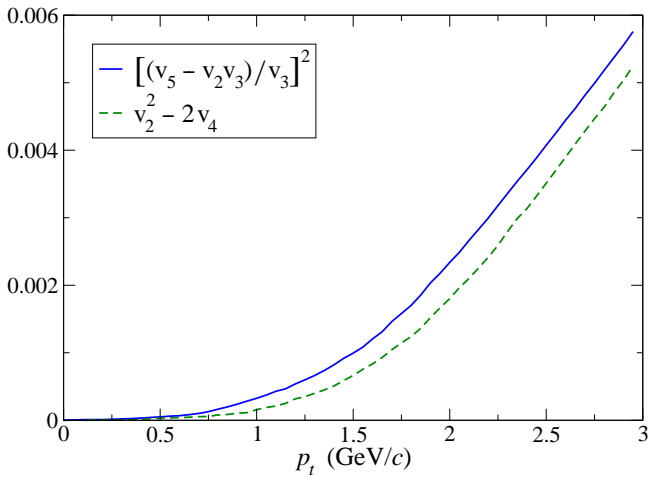}
\caption{\label{fig:vn_relations} Combinations of flow harmonics [full: squared lhs of Eq.~(\ref{vn_relation_1}); dashed: lhs of Eq.~(\ref{vn_relation_2})] for pions in a full computation of the Cooper--Frye integral with viscous corrections.}
\end{figure}

Figures~\ref{fig:delta_v2} and \ref{fig:delta_v3} show that the saddle-point calculation correctly approximates the correction arising from the additive dissipative term at freeze out. 
The displayed quantities are, however, not experimental observables and thus this particular result can only be of use for numerical simulations, in which the corrections can be turned on or off at will. 
In contrast, the combinations on the left-hand sides (lhs) of Eqs.~(\ref{vn_relation_1}) and~(\ref{vn_relation_2}) only involve measurable quantities. 
In Fig.~\ref{fig:vn_relations}, we show the squared lhs of Eq.~(\ref{vn_relation_1}) and the lhs of Eq.~(\ref{vn_relation_2}), computed within our exact freeze-out model with shear viscosity. 
From those equations, they should be equal, namely to the squared dissipative contribution to $v_2(p_t)$. 
The danger here is that these combinations of flow coefficients do not vanish when computed with the harmonics $v_n(p_t)$ obtained in ``exact'' calculations without dissipative correction---and accordingly they are about a factor of 2--3 larger than $[\mathcal{D}(p_t)_{}V_2]^2$. 
This is somewhat disappointing, yet we view the good agreement---which persists for other sets of parameters---between the two curves in Fig.~\ref{fig:vn_relations} as a hint that the displayed quantities open the possibility to pin down the effects of dissipation at decoupling, although we could not come up with a crisp mathematical argument to substantiate that statement. 

Now, in this paper, we have admittedly left aside a few phenomena, which could spoil the validity of our findings. 
Initial-state fluctuations are probably not hard to deal with, as their effect is to add analysis-method-dependent multiplicative coefficients~\cite{Gombeaud:2009ye,Teaney:2012ke}---related to the initial eccentricities, and which can be deduced from the study of integrated flow or from that of slow particles---in front of the flow harmonics in relations~(\ref{vn_relation_1}) or~(\ref{vn_relation_2}). 

A possibly more worrisome effect is that of rescatterings, if any, after the fluid-particle transition, which might blur the relations by contributing some more anisotropic flow. 
Again, we think this difficulty can be handled, first by exploiting particles that rescatter less and secondly by gauging the influence of hadronic collisions in transport models. 
Here we wish to mention an interesting possibility: by investigating particles with different cross sections, i.e.\ that decouple at different stages of the evolution, one could ideally hope to map---once the functional form of $\overline{\delta f}$ is known, although it admittedly depends on the particle type!---the temperature dependence of the transport coefficients in some region in the hadronic phase. 
In the future, we wish to investigate this idea within more realistic numerical simulations.

\begin{acknowledgments} 
  We thank Ulrich Heinz, Tetsufumi Hirano, Scott Pratt and Raimond Snellings for their questions and comments to preliminary presentations of this work, which hopefully helped us clarify some points of our message. 
\end{acknowledgments}

\appendix

\section{Details of the saddle-point calculations}
\label{app:calculations}

In this appendix, we give the main steps of the saddle-point calculations of the Cooper--Frye integral. 
For the sake of brevity, the $\mathsf{x}$ dependence of the velocity will often be dropped in the following. 

Let $\textbf{p}_t$ und $y$ be the transverse momentum and longitudinal rapidity of an emitted particle of mass $m$ in a fixed ``laboratory'' frame and $m_t$ its transverse mass. 
We denote by $\textbf{u}_t$ resp.\ $y_f$ the transverse component resp.\ longitudinal rapidity of the fluid four-velocity, whose timelike component is then given by $\sqrt{1+{\bf u}_t^2}\cosh y_f$, as follows from the normalisation $u_\mu u^\mu = 1$. 
Eventually, $\varphi$ resp.\ $\varphi_f$ stands for the azimuth of the particle transverse momentum resp.\ of the fluid transverse velocity with respect to some fixed axis. 

With these notations, one quickly finds
\begin{align}
\mathsf{p}\cdot\mathsf{u} &= m_t\sqrt{1+{\bf u}_t^2}\cosh(y-y_f) - \textbf{p}_t\cdot\textbf{u}_t \cr
 &= m_t\sqrt{1+u_t^2}\cosh(y-y_f) - p_tu_t\cos(\varphi-\varphi_f),\qquad \label{p.u_1}
\end{align}
with $p_t = |{\bf p}_t|$, $u_t = |{\bf u}_t|$. 

It is convenient to introduce the transverse rapidity $y_{f,t}$ of the fluid---defined through $u_t = \sinh y_{f,t}$---and that of the emitted particle, $y_t$, which obeys $m_t = m\cosh y_t$, $p_t=m\sinh y_t$. 
With their help, the inner product~(\ref{p.u_1}) becomes
\begin{align}
\label{p.u_2}
\mathsf{p}\cdot\mathsf{u} =\ &m\cosh y_t \cosh y_{f,t} \cosh(y-y_f) \cr
 & - m\sinh y_t\sinh y_{f,t}\cos(\varphi-\varphi_f). 
\end{align}
Minimising $\mathsf{p}\cdot\mathsf{u}$ with respect to $y_f$ or $\varphi_f$ is trivial and gives the conditions 
\begin{equation}
\label{conditions1}
y_f = y\quad\text{and}\quad\varphi_f = \varphi. 
\end{equation}
The former means that the time and longitudinal components (in a Cartesian coordinate system) of the particle four-momentum and the fluid four-velocity obey $p^z/p^0=u^z/u^0 (=\tanh y)$ at the point(s) of the freeze-out surface where $\mathsf{p}\cdot\mathsf{u}(\mathsf{x})/T$ is minimal. 
In turn, the equality $\varphi_f = \varphi$ means that the transverse components of the corresponding four-vectors are parallel at that (those) emission point(s). 

Straightforward computations yield the second derivatives of $\mathsf{p}\cdot\mathsf{u}(\mathsf{x})$ with respect to either $y_f$ or $\varphi_f$; at the minimum, their respective values are
\begin{subequations}
\label{2nd_derivatives1}
\begin{align}
\frac{\partial^2(\mathsf{p}\cdot\mathsf{u})}{\partial y_f^2}\bigg|_{\min} &= m\cosh y_t \cosh y_{f,t}, \\
\frac{\partial^2(\mathsf{p}\cdot\mathsf{u})}{\partial\varphi_f^2}\bigg|_{\min} &= m\sinh y_t \sinh y_{f,t}.
\end{align}
\end{subequations}

Under conditions~(\ref{conditions1}), the inner product of particle four-momentum and fluid four-velocity reads
\begin{align}
\mathsf{p}\cdot\mathsf{u} &= m\cosh y_t \cosh y_{f,t} - m\sinh y_t\sinh y_{f,t} \cr
 &= m\cosh(y_t - y_{f,t}), \label{p.u_min1}
\end{align}
which is clearly minimal when $y_t - y_{f,t}$ is smallest. 
Note that the first line can also be rewritten as
\begin{equation}
\mathsf{p}\cdot\mathsf{u} = m_t \cosh y_{f,t} - p_t\sinh y_{f,t}. \label{p.u_min1b}
\end{equation}

Instead of characterizing the transverse components of four-vectors---particle momentum or fluid velocity---through the azimuthal angle and transverse rapidity, one may adopt the same choice as in Ref.~\cite{Borghini:2005kd}. 
Fixing the four-momentum of the emitted particle---or actually, its transverse momentum ${\bf p}_t$---, one considers the components of the transverse fluid velocity parallel and orthogonal to ${\bf p}_t$, denoted respectively by $u_\parallel$ and $u_\perp$.
In that coordinate system, Eq.~(\ref{p.u_1}) reads
\begin{equation}
\label{p.u_3}
\mathsf{p}\cdot\mathsf{u} = m_t\sqrt{1+u_\parallel^2+u_\perp^2}\cosh(y-y_f) - p_tu_\parallel,
\end{equation}
which is obviously minimum when 
\begin{equation}
\label{conditions2}
y_f = y\quad\text{and}\quad u_\perp=0,
\end{equation}
equivalent to conditions~(\ref{conditions1}) and resulting in 
\begin{equation}
\label{p.u_min2}
\mathsf{p}\cdot\mathsf{u} = m_t\sqrt{1+u_\parallel^2} - p_t u_\parallel,
\end{equation}
whose minimum is reached when $v_\parallel\equiv u_\parallel/\sqrt{1+u_\parallel^2}$ is as close as possible to $p_t/m_t$---which naturally amounts to $y_t - y_{f,t}$ being smallest. 
The second derivatives of $\mathsf{p}\cdot\mathsf{u}(\mathsf{x})$ with respect to either $y_f$ or $u_\perp$ at the minimum are
\begin{subequations}
\label{2nd_derivatives2}
\begin{align}
\frac{\partial^2(\mathsf{p}\cdot\mathsf{u})}{\partial y_f^2}\bigg|_{\min} &= m_t\sqrt{1+u_\parallel^2}, \\
\frac{\partial^2(\mathsf{p}\cdot\mathsf{u})}{\partial u_\perp^2}\bigg|_{\min} &= \frac{m_t}{\sqrt{1+u_\parallel^2}}.
\end{align}
\end{subequations}

We now proceed with the minimisation of $\mathsf{p}\cdot\mathsf{u}(\mathsf{x})$ and discuss the distinction between slow and fast particles. 

\subsection{Slow particles}

If there is a point on the freeze-out hypersurface such that $y_{f,t}(\mathsf{x})=y_t$---which in the terminology introduced in Ref.~\cite{Borghini:2005kd} defines ``slow particles''---then it gives the minimum of $\mathsf{p}\cdot\mathsf{u}(\mathsf{x})$, which simply equals $m$ [see Eq.~(\ref{p.u_slow})]. 
As thus, this constitutes the saddle point $\mathsf{x}_{\mathrm{s.p.}}$ of the Cooper--Frye integral. 

One easily checks that conditions~(\ref{conditions1}) together with $y_{f,t}=y_t$ are equivalent to the relation $p^\mu = mu^\mu(\mathsf{x}_{\mathrm{s.p.}})$ at the saddle point(s). 
The second derivative of $\mathsf{p}\cdot\mathsf{u}(\mathsf{x})$ with respect to $y_{f,t}$ is trivially equal to $\mathsf{p}\cdot\mathsf{u}(\mathsf{x})$ itself.
Making the substitution $y_{f,t}=y_t$ in Eq.~(\ref{2nd_derivatives1}) yields for the non-vanishing second derivatives at the saddle point
\begin{subequations}
\label{2nd_derivatives_slow1}
\begin{align}
\frac{\partial^2(\mathsf{p}\cdot\mathsf{u})}{\partial y_f^2}\bigg|_{\min\!} &= m\cosh^2 y_t = \frac{m_t^2}{m}, \\
\frac{\partial^2(\mathsf{p}\cdot\mathsf{u})}{\partial\varphi_f^2}\bigg|_{\min\!} &= m\sinh^2 y_t = \frac{p_t^2}{m}, \\
\frac{\partial^2(\mathsf{p}\cdot\mathsf{u})}{\partial y_{f,t}^2}\bigg|_{\min\!} &= m \quad\text{for slow particles.}
\end{align}
\end{subequations}

Equivalently, in the ${\bf p}_t$-attached coordinate system with ($u_\parallel$, $u_\perp$) components, the identity $y_{f,t}=y_t$ becomes $v_\parallel=p_t/m_t$, and the second derivatives of $\mathsf{p}\cdot\mathsf{u}(\mathsf{x})$ at the minimum read
\begin{subequations}
\label{2nd_derivatives_slow2}
\begin{align}
\frac{\partial^2(\mathsf{p}\cdot\mathsf{u})}{\partial y_f^2}\bigg|_{\min\!} &= \frac{m_t^2}{m}, \\
\frac{\partial^2(\mathsf{p}\cdot\mathsf{u})}{\partial u_\perp^2}\bigg|_{\min\!} &= m, \\
\frac{\partial^2(\mathsf{p}\cdot\mathsf{u})}{\partial u_\parallel^2}\bigg|_{\min\!} &= \frac{m^3}{m_t^2} \quad\text{for slow particles.}
\end{align}
\end{subequations}
In the saddle-point calculation, these derivatives, divided by $T$, become the inverse widths of Gaussians which are integrated over. 

To ensure the validity of the saddle-point approximation, the higher terms in the Taylor expansion of $\mathsf{p}\cdot\mathsf{u}(\mathsf{x})/T$ should be negligible compared to the quadratic ones. 
Considering for instance the derivatives with respect to the transverse rapidity, the odd ones vanish at the saddle point while the even ones all are equal to $m/T$, as shown by Eq.~(\ref{p.u_min1}).
Fixing momentarily $y_f$ and $\varphi_f$ to their saddle-point values, one thus has
\[
\frac{\mathsf{p}\cdot\mathsf{u}(\mathsf{x})}{T} \sim \frac{m}{T} + \frac{m}{T}\frac{(y_{f,t}-y_t)^2}{2} + \frac{m}{T}\frac{(y_{f,t}-y_t)^4}{4!} + \cdots
\]
The quadratic term in this expression is at most unity for values of $y_{f,t}-y_t\lesssim\sqrt{T/m}$. 
The quartic term is then much smaller than the quadratic one provided $m\gg T$. 
This strong inequality constitutes a second condition---besides that regarding their transverse velocity---to be fulfilled by slow particles for the saddle-point calculation to hold.

\subsection{Fast particles}

For ``fast particles'', defined as those whose transverse velocity is larger than the maximal transverse velocity $u_{\max}(y_f,\varphi_f)$ reached by the fluid flowing in the same direction, ${\sf p}\cdot{\sf u}(\textsf{x})$ is minimal when the fluid transverse velocity takes its maximum value along that direction, namely $y_{f,t}^{\max}(y_f,\varphi_f)=\ln[u_{\max}(y_f,\varphi_f)+u^0_{\max}(y_f,\varphi_f)]$, where we have defined $u_{\max}^0(y_f,\varphi_f) = \sqrt{1+u_{\max}(y_f,\varphi_f)^2}$. 
This gives the value of the product~(\ref{p.u_2}) at the corresponding point on the freeze-out hypersurface, namely
\begin{equation}
\label{p.u_fast_app}
{\sf p}\cdot{\sf u}(\mathsf{x}_{\mathrm{s.p.}}) = m\cosh y_t\, u_{\max}^0(y,\varphi) - m\sinh y_t\, u_{\max}(y,\varphi), 
\end{equation}
which is equivalent to Eq.~(\ref{p.u_fast}). 
At that saddle point, one also finds the first derivative
\begin{subequations}
\begin{equation}
\label{1st_derivative_fast1}
\frac{\partial({\sf p}\cdot{\sf u})}{\partial y_{f,t}}\bigg|_{\min\!} = m_t u_{\max}(y,\varphi) - p_t u_{\max}^0(y,\varphi). 
\end{equation}
or equivalently, in the ${\bf p}_t$-attached coordinate system of Ref.~\cite{Borghini:2005kd},
\begin{equation}
\label{1st_derivative_fast2}
\frac{\partial({\sf p}\cdot{\sf u})}{\partial u_\parallel}\bigg|_{\min\!} = m_t v_{\max}(y,\varphi) - p_t,
\end{equation}
\end{subequations}
with $v_{\max}(y,\varphi) \equiv u_{\max}(y,\varphi)/u_{\max}^0(y,\varphi)$. 
The latter expression shows at once that this derivative is negative, since $p_t/m_t>v_{\max}(y,\varphi)$. 
As for slow particles, the other first two derivatives vanish. 

In turn the non-vanishing second derivatives are
\begin{subequations}
\label{2nd_derivatives_fast}
\begin{align}
\label{2nd_derivatives_fast1}
\frac{\partial^2({\sf p}\cdot{\sf u})}{\partial y_f^2}\bigg|_{\min\!} &= m_t u^0_{\max}(y,\varphi), \\
\frac{\partial^2({\sf p}\cdot{\sf u})}{\partial\varphi_f^2}\bigg|_{\min\!} &= p_tu_{\max}(y,\varphi), \\
\frac{\partial^2({\sf p}\cdot{\sf u})}{\partial y_{f,t}^2}\bigg|_{\min\!} &= m_t u_{\max}^0(y,\varphi) - p_t u_{\max}(y,\varphi) \\
&\quad\text{ for fast particles.} \nonumber
\end{align}
\end{subequations}
The second derivative with respect to $y_{f,t}$ is actually irrelevant for the saddle-point calculation, since the corresponding first derivative does not vanish and thus is the leading term of the approximation. 
It is, however, important to determine the region of validity of the approximation. 
Writing, with $y_f$ and $\varphi_f$ fixed to their saddle-point values (note that $y_{f,t}-y_t \leq 0$, so that the linear term is actually positive despite the negative derivative)
\begin{align*}
\frac{\mathsf{p}\cdot\mathsf{u}(\mathsf{x})}{T} \sim &\ \frac{m_t u^0_{\max}-p_t u_{\max}}{T} \\
 &\quad- \frac{p_t u^0_{\max} - m_t u_{\max}}{T}(y_{f,t}-y_t) \\
 &\quad+ \frac{m_t u^0_{\max} - p_t u_{\max}}{T}\frac{(y_{f,t}-y_t)^2}{2} + \cdots, 
\end{align*}
one finds that the quadratic term is negligible compared to the linear one provided
\begin{equation}
\label{criterion_fast}
\frac{(m_t u_{\max}-p_t u^0_{\max})^2}{m_t u^0_{\max}-p_t u_{\max}} \gg T. 
\end{equation}
This is the criterion---actually more stringent than the condition $p_t>m_t v_{\max}$---which gives the validity region of the saddle-point approximation for fast particles.

\section{Conformal second-order correction to the phase space distributions}
\label{app:Teaney-Yan_eqs}

To make this paper more self-contained, we gather hereafter the expression of the conformal second-order correction to the phase space distribution as computed within the relaxation time approximation by Teaney and Yan~\cite{Teaney:2013gca}, to whose paper we refer the reader for more details. 
Note that we give here the relative correction $\overline{\delta f}^{(2)}$, while equations~(2.30) and (2.31) of Ref.~\cite{Teaney:2013gca} relate to the absolute correction $\delta f^{(2)} = \overline{\delta f}^{(2)}\times f_0$. 
Accordingly, the scalar quantities $\bar{\chi}_{ip}$, $\bar{\xi}_{jp}$ we discuss below differ from their counterparts $\chi_{ip}$, $\xi_{jp}$ in Ref.~\cite{Teaney:2013gca}. 

We denote by $\pi^{\mu\nu}$ the dissipative stress tensor up to second order in the conformal case, that is, including not only the first-order correction proportional to $\eta$, but also the second-order terms involving the relaxation time $\tau_\pi$ and the ``usual'' coefficients $\lambda_1$, $\lambda_2$, $\lambda_3$~\cite{Baier:2007ix}.
$\Delta^{\mu\nu}$ is the projector on the local rest frame, i.e. orthogonal to the fluid four-velocity, and $\nabla^\mu\equiv \Delta^{\mu\nu}\partial_\nu$ denotes the spatial derivatives in that frame. 

With these notations, the conformal second-order relative correction to the phase space distribution at freeze out reads
\begin{widetext}
\begin{align}
\label{df2_Teaney-Yan}
\overline{\delta f}^{(2)} =
   \frac{\bar{\chi}_{1p}}{\eta^2}\,\frac{p^{\mu_1}p^{\mu_2}p^{\mu_3}p^{\mu_4} }{T^6}\,
     \pi_{\langle\mu_1\mu_2} \pi_{\mu_3 \mu_4\rangle} 
   &+  \frac{\bar{\chi}_{2p}} {\eta} \,\frac{p^{\mu_1} p^{\mu_2} p^{\mu_3}}{T^5} 
     \left[   \frac{6}{T}\,\pi_{\langle\mu_1\mu_2} \nabla_{\mu_3\rangle}T - \nabla_{\langle\mu_1} \pi_{\mu_2 \mu_3\rangle} \right] \cr
  &+ \frac{\bar{\xi}_{1p}}{\eta^2} \, \frac{p^{\mu_2} p^{\mu_1}  }{T^4} \pi^\lambda_{\;\langle\mu_2}\pi_{\mu_1\rangle\lambda}
   + \frac{\bar{\xi}_{2p}}{\eta} \frac{p^{\mu_2} p^{\mu_1}}{T^3} \left[ \pi_{\mu_2\mu_1} + \eta\nabla_{\langle\mu_2}u_{\mu_1\rangle} \right]    \cr 
  &+ \frac{\bar{\xi}_{3p}}{\eta} \frac{p^{\mu_2}}{T^3} \left[ \Delta_{\mu_2\lambda_2} \partial_{\lambda_1} \pi^{\lambda_1 \lambda_2} \right] 
   +  \frac{\bar{\xi}_{4p}}{T^2 \eta^2} \pi^{\mu\nu}\pi_{\mu\nu}\,, 
\end{align}
\end{widetext}
where the angular brackets denote the construction of a traceless symmetric tensor orthogonal to the fluid four-velocity.
Together with a third function $\bar{\chi}_{0p}$ which will appear hereafter, $\bar{\chi}_{1p}$ and $\bar{\chi}_{2p}$ are related to the equilibrium distribution (for $\bar{\chi}_{1p}$) and to the relaxation time of the corresponding approximation, yet their precise expressions will not be needed. 
In equation~(\ref{df2_Teaney-Yan}), the four scalar functions $\bar{\xi}_{1p}$, $\bar{\xi}_{2p}$, $\bar{\xi}_{3p}$, $\bar{\xi}_{4p}$ are linearly related to $\bar{\chi}_{0p}$, $\bar{\chi}_{1p}$, and $\bar{\chi}_{2p}$ through
\begin{subequations}
\label{xieq}
\begin{align}
   \bar{\xi}_{1p} &= \bar{\chi}_{1p} \, \frac{4\,\textbf{p}_{\rm LR}^2}{7\,T^2} - \frac{\bar{\chi}_{2p}\,E_{\textbf{p},\rm LR}}{\eta\tau_\pi T} (\eta\tau_\pi+ \lambda_1) \, ,  \\
   \bar{\xi}_{2p} &=  \frac{\bar{\chi}_{2p}\,E_{\textbf{p},\rm LR}}{T^2\tau_\pi} - \bar{\chi}_{0p}   \, , \\
   \bar{\xi}_{3p} &=  -\bar{\chi}_{2p} \frac{2\,\textbf{p}_{\rm LR}^2}{5\,T^2}  + 2 \bar{\chi}_{0p} \frac{\eta}{s} \frac{E_{\textbf{p},\rm LR}}{T} -  a_{\textsf{p}_*}\, ,  \\
   \bar{\xi}_{4p} &=  \bar{\chi}_{1p} \frac{2\,\textbf{p}_{\rm LR}^4}{15\,T^4} - 
   \bar{\chi}_{2p} \frac{E_{\textbf{p},\rm LR}\, \textbf{p}_{\rm LR}^2}{3\,T^3} \cr
 &\qquad\qquad\ \quad\,- \bar{\chi}_{0p} \frac{\eta}{s} \frac{E_{\textbf{p},\rm LR}^2}{T^2} c_s^2 - a_{E_*}  \frac{E_{\textbf{p},\rm LR}}{T} \, ,\qquad\label{xi4}
\end{align}
\end{subequations}
with $E_{\textbf{p},\rm LR}$ and $\textbf{p}_{\rm LR}$ the particle energy and momentum in the local rest frame, and $a_{E_*}$, $a_{\textsf{p}_*}$ two coefficients discussed in Ref~\cite{Teaney:2013gca} that play no role here.

\section{Anisotropic flow modification from the shear viscous correction at freeze out}
\label{app:df1_shear}

In this appendix, we illustrate the passage from a given dissipative correction $\overline{\delta f}$ to the phase space distribution to the corresponding function $\mathcal{D}(p_t)$ in Eqs.~(\ref{vn(pt)}) on an example.

Within the framework of Grad's prescription, the relative correction to the phase space distribution due to shear viscous effects reads
\begin{align}
\label{df_shear}
\overline{\delta f}^{(1)}_{\mathrm{shear}} &= 
  \frac{1}{2[e(\mathsf{x})+\mathcal{P}(\mathsf{x})]_{}T(\mathsf{x})^2}_{} \pi^{\mu\nu}_{\mathrm{shear}}(\mathsf{x})_{} p_\mu p_\nu \cr
  &\equiv C_{\mathrm{shear}}(\mathsf{x})_{}\pi^{\mu\nu}_{\mathrm{shear}}(\mathsf{x})_{} p_\mu p_\nu,
\end{align}
with $e$ resp.\ $\mathcal{P}$ the energy density resp.\ pressure and $\pi^{\mu\nu}_{\mathrm{shear}}$ the (traceless) shear viscous part of the stress dissipative tensor. 
The latter can be recast as $\pi^{\mu\nu}_{\mathrm{shear}} = \eta_{}\nabla^{\langle\mu}u^{\nu\rangle}$, with $\eta$ the shear viscosity.

To handle the product $\pi^{\mu\nu}_{\mathrm{shear}}p_\mu p_\nu$, one can again use the Landau matching condition, to cancel out the component of the particle four-momentum $p^\mu$ along the flow velocity $u^\mu$.
Introducing $q^\mu \equiv p^\mu - (p^0/u^0)u^\mu$, one thus obtains $\pi^{\mu\nu}_{\mathrm{shear}}p_\mu p_\nu = \pi^{\mu\nu}_{\mathrm{shear}}q_\mu q_\nu$. 

Let us make a few assumptions on the velocity profile at freeze out. 
We thus assume that it is approximately radial and neglect the azimuthal modulation of its gradient---which amounts to leaving aside higher-order terms in some small Fourier coefficients, as already mentioned above Eqs.~(\ref{vn(pt)}). 
Under those hypotheses, only the radial component $p_t - m_t v_{\max}(\varphi)$, with $v_{\max}(\varphi)\equiv u_{\max}(\varphi)/u^0_{\max}(\varphi)$, of $\textsf{q}$ plays a role, and one finds at the saddle point $\mathsf{x}_{\textrm{s.p.}}$ corresponding to a given four-momentum $\textsf{p}$
\[
\overline{\delta f}^{(1)}_{\mathrm{shear}} = 
  C_{\mathrm{shear}}(\mathsf{x}_{\textrm{s.p.}})_{}\eta_{}\nabla^{\langle r}u^{r\rangle}(\mathsf{x}_{\textrm{s.p.}})
  \big[p_t - m_t v_{\max}(\varphi)\big]^2,
\]
where the superscript $r$ denotes the radial direction. 

Let $C'_{\mathrm{shear}} \equiv C_{\mathrm{shear}}(\mathsf{x}_{\textrm{s.p.}})_{}\eta\,\nabla^{\langle r}u^{r\rangle}(\mathsf{x}_{\textrm{s.p.}})$. 
Using the Fourier expansion~(\ref{u_max(phi)_expansion}) to expand $v_{\max}(\varphi)$, one finds after some algebra
\begin{widetext}
\begin{equation}
\label{drosophile}
1 + \overline{\delta f}^{(1)}_{\mathrm{shear}} \simeq \Big[ 1 + C'_{\mathrm{shear}} \big(p_t - m_t \bar{v}_{\max}\big)^2\Big]
 \left[ 1 - 
  \frac{2_{}C'_{\mathrm{shear}} \big(p_t - m_t \bar{v}_{\max}\big)}{1+C'_{\mathrm{shear}} \big(p_t - m_t \bar{v}_{\max}\big)^2}
  \frac{m_t\bar{v}_{\max}}{1+\bar{u}_{\max}^2}\sum_{n\geq 1}2V_n\cos n(\varphi-\Psi_n) \right].
\end{equation}
\end{widetext}
The first factor is independent of azimuth, and thus does not contribute to the flow coefficients $v_n(p_t)$. 
In turn, the second factor yields $\mathcal{D}(p_t)$:
\begin{equation}
\label{D(pt)_shear}
\mathcal{D}(p_t)_{\textrm{shear}} = 
  \frac{2_{}C'_{\mathrm{shear}} \big(p_t - m_t \bar{v}_{\max}\big)}{1+C'_{\mathrm{shear}} \big(p_t - m_t \bar{v}_{\max}\big)^2}
  \frac{m_t\bar{v}_{\max}}{1+\bar{u}_{\max}^2}.
\end{equation}
The terms in $\mathcal{I}(p_t)$ in Eqs.~(\ref{vn(pt)}) come from the exponential in Eq.~(\ref{spectrum_fast}), which eventually multiplies Eq.~(\ref{drosophile}).

$C'_{\mathrm{shear}} \big(p_t - m_t \bar{v}_{\max}\big)^2$ should be significantly smaller than 1, so that viscous corrections remain small. 
The dependence of $\mathcal{D}(p_t)$ on transverse momentum is thus actually given by the numerator, and is approximately quadratic.

\end{document}